\documentclass[conference]{IEEEtran}
\usepackage{amsmath}
\newcommand{\code}[1]{{\small \tt #1}}
\ifCLASSINFOpdf
  \usepackage[pdftex]{graphicx}
\else
\fi

\begin{document}
%\title{Man-in-the-Middle Attack on Bitcoin over Tor}
% reference to: https://blog.torproject.org/blog/bittorrent-over-tor-isnt-good-idea
\title{Bitcoin over Tor isn't a good idea}

\author{
\IEEEauthorblockN{Alex Biryukov}
\IEEEauthorblockA{University of Luxembourg\\
Email: alex.biryukov@uni.lu} 
\and
\IEEEauthorblockN{Ivan Pustogarov}
\IEEEauthorblockA{University of Luxembourg\\
Email: ivan.pustogarov@uni.lu}
}


%Telephone: (800) 555--1212, Fax: (888) 555--1212}


\maketitle

\begin{abstract}

Bitcoin is a decentralized P2P digital currency in which coins are
generated by a distributed set of miners and transaction are broadcasted
via a peer-to-peer network. While Bitcoin provides some level of anonymity
(or rather pseudonymity) by encouraging the users to have any number of random-looking 
Bitcoin addresses,
recent research shows that this level of anonymity is rather low. This encourages
users to connect to the Bitcoin network through anonymizers like Tor and motivates 
development of default Tor functionality for popular mobile SPV clients.
In this paper we show that combining Tor and Bitcoin 
creates an attack vector for the deterministic and stealthy
man-in-the-middle attacks. 
A low-resource attacker can gain full control of information flows between all
users who chose to use Bitcoin over Tor.
In particular the attacker can link together user's transactions regardless of pseudonyms used, control which Bitcoin blocks and transactions are relayed to the user and can \
delay or discard user's transactions and blocks. In collusion with a powerful miner
double-spending attacks become possible and a totally virtual Bitcoin reality can be created for such set of users.
Moreover, we show how an attacker can fingerprint users and then recognize them 
and learn their IP address when they decide to connect to the Bitcoin network directly.
\end{abstract}


% \fi
\IEEEpeerreviewmaketitle

\section{Introduction}
\label{sec:introduction}
Bitcoin is a decentralized virtual currency and a P2P payment system in which coins are generated
by miners and double spending is prevented by that each peer keeps a local
copy of the constantly growing public ledger of all the previous transactions.
Though the original Bitcoin paper states that privacy in such a system may
still be maintained, the recent findings disprove this. Anonymity and privacy of the plain Bitcoin protocol is also not claimed by the Bitcoin developers.

There are two independent problems:
a) ability of the attacker to link transactions to the IP address of the user~\cite{deanonbitcoin},~\cite{koshyanalysis},~\cite{dankamtalk} by studying connectivity and traffic of the peers and b) linkability of the user's pseudonyms and transactions
in the public ledger achieved via graph and transaction flow analysis~\cite{ron2013quantitative,meiklejohn}.
At the same time as Bitcoin increases its user base and moves from mining and hoarding 
to the actual use as a currency and payment protocol in various on-line applications
there is a growing demand in more privacy among the Bitcoin users. 
While one could use a Bitcoin mixing service\footnote{This is always a matter of trust of the service operator.} 
to break connections in the transaction graph, IP address leakage is still possible. 
Bitcoin developers recommend to use third party anonymization tools like Tor or 
VPNs to solve this problem.

Seeing this shortcomming of Bitcoin some alternative currencies like Anoncoin, BitTor, Torcoin, Stealthcoin  and others offer native support for Tor. There are also several other use cases for Tor in the Bitcoin ecosystem.  For mobile payments it is of interest to use so called SPV (simple payment verification) clients which cannot afford to hold the full 20 Gbyte blockchain ledger. Such feature was already foreseen in the original Bitcoin whitepaper, see Section~8 of~\cite{nakamoto2012bitcoin}.  Since such popular clients (around 1 Million expected userbase~\cite{hearn-bitcoinj}) are vulnerable to spoofing attacks which may result in double-spending, the current trend is to bundle them with Tor by default to avoid spoofing and  man-in-the-middle attacks~\cite{bitcoinj,Electrum}.  Tor can also be a solution for services and online shops that want to prevent DoS attacks against their public IP. Finaly Tor is seen as a countermeasure if Internet neutrality towards Bitcoin will start to erode~\cite{bitcoin-neutrality}. 

Tor is not a panacea however and not all applications are anonymized
equally well when combined with Tor.
The biggest effort has been made so far on improving protection of
the HTTP(S) protocol on top of Tor. Other protocols are not researched
that well. There were several documented cases
when application level leaked crucial user-identifying information
\cite{bittorrenttor}, \cite{flashtor}.
Moreover, there is only limited number of applications
which are studied well enough to be considered safe to use with Tor
\cite{torsocks}. 

This paper contains two main contributions: first we show that 
using Bitcoin through Tor not only provides
limited level of anonymity but also exposes the user to man-in-the middle attacks
in which an attacker  controls which Bitcoin blocks and
transactions the users is aware of. Moreover in collusion with a powerful miner
double-spending becomes possible and a totally virtual Bitcoin reality may be created for such users 
(at least for a brief period of time).

The second main contribution is a fingerprinting technique for Bitcoin users by setting an ``address cookie'' on the user's computer. This can be used to correlate the
same user across different sessions, even if he uses Tor, hidden-services or multiple proxies. If the user later decides to connect to the Bitcoin network directly the cookie would be still present and would reveal his IP address. A small set of Sybil nodes (about a 100 attacker's nodes) is sufficient to keep the cookies fresh on all the Bitcoin peers (including clients behind NATs).

%The attacks presented in this paper are made possible by exploiting
%Bitcoin's reputation based DoS protection.
%Whenever a peer receives a malformed
%message, it increases the penalty score of the IP address
%from which the message came (if a client uses Tor, then the
%message will obviously come from one of the Tor exit nodes).
%When this score exceeds 100, the sender’s IP is banned for 24
%hours.

The man-in-the-middle attack exploits a Bitcoin built-in reputation based
DoS protection and the attacker is able to force specific Bitcoin peers to ban Tor Exit nodes of her choice. Combining it with some peculiarities of how
Tor handles data streams a stealthy and low-resource 
attacker with just 1-3\% of overall Tor Exit bandwidth capacity and 1000-1500 cheap lightweight Bitcoin peers (for example, a small Botnet) can force \textit{all} Bitcoin Tor traffic to go either through her Exit nodes or through her peers. This opens numerous attack vectors. First
it simplifies a traffic correlation attack since the attacker controls
one end of the communication. Second, the attacker can can glue
together different Bitcoin addresses (pseudonyms) of the same user. Third, it opens possibilities of double spending attacks for the mobile SPV clients, those which it was supposed to protect from such attacks. The estimated cost of the attack is below 2500 USD per month.  Issues described in this paper were experimentally verified, by tracking our own clients in the real Bitcoin and Tor networks. We also notified Tor and Bitcoin developers about these vulnerabilities of their protocols.

The rest of the paper is organized as follows. In section \ref{sec:background} we
provide information on Bitcoin and Tor internals required for understanding the attacks. In section \ref{sec:getinthemiddle} we describe how an attacker
can get in the middle between Bitcoin clients and Bitcon network, effectively isolating the client from the rest of the Bitcoin P2P network.
We also show that Bitcoin peers available as Tor hidden services may not solve the problem.
In section~\ref{sec:fingerprinting} we show how a user can be fingerprinted and his activity
linked across different sessions. Section~\ref{sec:portspoof} describes how the
attacker can increase the probability that a user connecting to the Bitcon network
directly (i.g. without using Tor) will choose her peers.
In section~\ref{sec:delays} we analyze the man-in-the-middle attack 
and estimate connection delays experienced by
the user and check for which malicious Exit bandwidth and number
of malicious peers the attack becomes unnoticeable to the user.
%In section \ref{sec:attacks} we describe attacks which become possible.
Section \ref{sec:costsandrate} calculates the costs of the attack. In section \ref{sec:countermeasures} we describe several possible
countermeasures. 
%Section \ref{sec:conclusion} concludes the paper.

\section{Background}
\label{sec:background}
In this section we provide details of the inner working of Tor and
Bitcoin protocols. Many of these details
were obtained by an analysis of the
corresponding source code. This is especially true for Bitcoin for which 
there exists no official documentation except for the original white paper~\cite{nakamoto2012bitcoin} and Bitcoin Wiki\cite{bitcoin-wiki}.

\subsection{Bitcoin}
Bitcoin is a decentralized virtual currency and a payment system based on
cryptography and a peer-to-peer network.
Its main components are transactions and blocks. Blocks are created
by Bitcoin miners by solving cryptographic puzzles of controlled hardness (called
\textit{proofs of work}). The proof of work consists of finding a cryptographic hash 
value for a block of transactions which starts with a certain number of leading zero bits (32 when Bitcoin was first proposed, 67 zero bits at present). 
With each solved block a miner creates and earns 25 new Bitcoins.
Hash of the previous block is included into the
new block, which results in a chain of blocks or \textit{blockchain}.
The difficulty of the cryptographic puzzles is adjusted automatically by the network
so that the network generates one block every 10 minutes on the average.
Payers and payees of the system are identified by Bitcoin addresses
which are base58-encoded hashes of their public keys. Money transfers from
one Bitcoin address to another are done by creating a signed transaction and broadcasting it to the P2P network.
Transactions are included into blocks by miners; once
a transaction is buried under a sufficient number of blocks,
it becomes computationally impractical to double spend
coins in this transaction.

Bitcoin is a peer-to-peer system where each peer is supposed to
keep its copy of the blockchain, which plays a role of a public ledger. 
Whenever a block or a transaction
is generated by a peer, it is broadcasted to other peers in the
network. Upon receipt and verification of the block's proof of work 
the peer updates his copy of the blockchain.
Bitcoin software does not explicitly divide its functionality between
clients and servers, however Bitcoin peers can be grouped
into those which accept incoming connections (\textit{servers}) and
those which don't (\textit{clients}), i.e. peers behind network address translation (NAT) or firewalls.
Bitcoin users connecting to the Bitcoin network through Tor or VPN obviously also
do not accept incoming connections.

At the time of writing there are about 7,000 reachable Bitcoin servers and an estimated number of 100,000 clients.
By default Bitcoin peers (both clients and servers) try to maintain
8 outgoing connections to other peers in the network. If any of the
8 outgoing connections drop, a Bitcoin peer tries to replace them with the 
new connections. Using our terminology, a Bitcoin client  can only establish
a connection to a Bitcoin server. We call servers to which a client established
connection \textit{entry nodes} of this client.
By default a server can accept up to 117 incoming connections.
If this limit is reached all new connections are dropped.

\subsubsection{Bitcoin anti-DoS protection}
\label{sssec:ados}
As an anti-DoS protection, Bitcoin peers implement a reputation-based protocol
with each node keeping a penalty score for every other Bitcoin peer (identified 
by its IP address). Whenever a malformed message is sent to
the node, the latter increases the penalty score of the sender and bans the
``misbehaving'' IP address for 24 hours when the penalty reaches the value of 100.

\subsubsection{Bitcoin peers as Tor hidden services}
\label{sssec:ths}
Tor hidden services (see section \ref{sssec:torhs}) are service-agnostic
in the sense that any TCP-based service can be made available as a Tor hidden
service. This is used by Bitcoin which recognizes three types of addresses:
IPv4, IPv6, and OnionCat~\cite{onioncat}. Onioncat address format is a way to
represent an onion address as an IPv6 address: the first 6 bytes of an OnionCat
address are fixed and set to \code{FD87:D87E:EB43} and the other 10 bytes are
the hex version of the onion address (i.e. base32 decoded onion address after
removing the ``.onion'' part).

\subsubsection{Bitcion peer discovery and bootstrapping}
\label{sssec:bcboot}
Bitcoin implements several mechanisms for peer discovery and bootstrapping.
First, each Bitcoin peer keeps a database of IP addresses of peers previously
seen in the network. This database survives between Bitcoin client restarts.
This is done by dumping the database to the hard drive every 15 minutes and
on exit (as we will see later this facilitates setting a cookie on the user's computer).
Bitcoin peers periodically broadcast their addresses in the network.
In addition peers can request addresses from each other using \code{GETADDR}
messages and unsolicitely advertise addresses using \code{ADDR} messages.

If Tor is not used, when a Bitcoin clients starts, it first tries to populate
its address database by resolving 6 hard-coded hostnames\footnote{At
time time of this writing one of these hostnames constantly failed to resolve into
any IP address.}. If Tor is used, Bitcoin does not explicitely ask Tor to
resolve\footnote{When applications communicate with Tor they can either ask Tor to establish a connection to a hostname by sending a \code{CONNECT} command
or to resolve a hostname by sending a \code{RESOLVE} command.} them
but rather asks it to establish connections to these hostnames. 

If Tor is not used, the addresses for outgoing connections are taken from
the addresses database only. In case Tor is used, every second connection
is established to a DNS hostname. These DNS hostnames are called "oneshots"
and once the client establishes a connection to such a hostname it requests
a bunch of addresses form it and then disconnects and never tries to connect
to it again.
As a fallback if no addresses can be found at all, after 60 seconds of running
the Bitcoin client uses a list of 600 hard-coded bitcoin addresses.

Bitcoin nodes recognize three types of addresses: IPv4, IPv6,
and OnionCat~\cite{onioncat}. For each type of addresses the peer maintains a state variable
indicating if the bitcoin node is capable of using such address type.
These state variables become important when using Tor: the only address type which
is accepted from other peers is OnionCat type.
Curiously, this results in that all IPv4 and IPv6 addresses obtained
from oneshots are dropped and the client uses its original database.
The opposite case also holds: if Tor is not used, onion addresses are 
not stored in the address database.

Finally each address is accompanied by a timestamp which determines its
freshness.

\subsubsection{Choosing outgoing connections}
\label{ssec:buckets}
For each address in the addresses database, a Bitcoin peer maintains statistics 
which among other things includes when the address was last seen in the network,
if a connection to this address was ever established before, and timestamp
of such connection.
All addresses in the database are distributed between so called buckets.
There are 256 buckets for ``new'' addresses (addresses to which the bitcoin
client has never established a connection) and 64 for ``tried'' addresses (addresses
to which there was at least to one successful connection). Each bucket can
have at most 64 entries (which means that there can be 
at most 20480 addresses in the database).
When a peer establishes outgoing connections, it
chooses an address from ``tried'' buckets with probability $p = 0.9-0.1n$, where
$n$ is the number of already established outgoing connections. If an address
is advertised frequently enough it can be put into up to 4 different ``new'' buckets.
This obviously increases its chances to be selected by a user and to be transferred
to a ``tried'' bucket.

\subsection{Tor}
Tor is the most popular low-latency anonymity network which at
the time of this writing comprised of 6000-7000 routers with an estimated number of
daily users exceeding 500,000 (not counting the botnet-infected nodes). 
Tor is based on ideas of onion routing and telescoping path-building design.
When a user wants to connect to an Internet server while keeping his IP
address in secret from the server he chooses the path consisting of three Tor relays (called \textit{Guard}, \textit{Middle} and \textit{Exit}),
builds a \textit{circuit} and negotiates symmetric key with each one
of them using the telescoping technique. Before sending a message to the server,
the user encrypts it
using the negotiated keys. While the message travels along the circuit,
each relay strips off its layer of encryption. In this way the message arrives
at the final destination in its original form and each party knows
only the previous and the next hop.

Tor tries hard to achieve low traffic
latency to provide a good user experience, thus sacrificing some anonymity for
performance. To keep latency low and network throughput high, Tor relays do
not delay incoming messages and do not use padding. This makes Tor susceptible
to traffic confirmation attacks: if an attacker is able to sniff both ends of the
communication, she is able to confirm that a user communicated with the server.
If the first hop of a circuit is chosen at random then the probability that
a malicious node will be chosen as the first hop  (and thus will know the IP address 
of the user) converges to one with the number of circuits.
Due to this, each user has a set of three\footnote{Will be reduced down to one Guard
per user in the next Tor update\cite{Tor-relay-early}.} \textit{Guard nodes}.
When a user builds a circuit the first hop is chosen from the set of trusted Guard nodes.

The list of all Tor relays is assembled and distributed in the so called \textit{consensus}
document by nine trusted Tor authorities. For the purpose of traffic balancing the
bandwidth of each relay is measured and reported. A user chooses relays
for his circuits proportional to relay's bandwidth listed in the consensus. Each relay in the consensus
is identified by his fingerprint (or ID) which is the SHA-1 hash of its public key.

\subsubsection{Tor stream timeout policy}
\label{sssec:tortimeoutes}
Tor provides a SOCKS interface for application willing to connect to the Internet
anonymously. Each connection to the SOCKS port by an application is called
a \textit{stream}. For each new stream Tor tries to attach it either to an
existing circuit or to a newly built one. It then sends a \code{BEGIN} cell down
the circuit to the corresponding Exit node asking it to establish a connection to
the server requested by the application. In order to improve user's quality of 
service,  if Tor does not receive a reply from the Exit node within
10 or 15 seconds\footnote{Tor waits for 10 seconds for the first two
attempt and 15 seconds for the subsequent attempts.}, it drops the circuit and
tries another one. If none of the circuits worked for the stream during 2 minutes,
Tor gives up on it and sends a SOCKS general failure error message.

\subsubsection{Tor Exit policy}
In order to access a Web resource anonymously through a Tor circuit,
the Exit relay (the final relay in the circuit) should allow establishing
connections outside the Tor network. This makes Exit relay operators open
to numerous abuses. In order to make their life easier, Tor allows to
specify an Exit Policy, a list of IP addresses and ports to which the Exit
node is willing to establish connections and which destination are prohibited.
When a client establishes a circuit, he chooses only those Exit nodes which
allow connections to the corresponding destination.

\subsubsection{Tor Hidden Services}
\label{sssec:torhs}
Tor is mostly known for its ability to provide anonymity for clients accessing Internet services.
Tor Hidden Services are a less known feature of Tor which enables responder anonymity:
a service can be contacted by clients without revealing its physical location. In order
to achieve this a client and the hidden service choose at random and connect to a Tor
relay (\textit{rendezvous point}) and forward all the data through it.
In more detail:
\begin{enumerate}
  \item The hidden service generates a public key and chooses at random a small number of Tor relays (typically three)
        which become its \textit{introduction points}. The service maintains permanent connection
        to these relays.
  \item It then generates an \textit{HS descriptor} which contains the public key and the list
        of introduction points, and
  \item Publishes it at 6 different Tor relays having \code{HSDir}
        flag.\footnote{\code{HSDir} flag is assigned by Tor authorities to relays  which
        wish to be a part of a distributed database
        to store descriptors of Tor hidden services. A relay should be running for at least 25 hours
        to get this flag.} These are called \textit{responsible HS directories}. The choice of responsible HS directories
        is deterministic and depends on the hash of the hidden service's public key and current day.
  \item Introduction points are instructed by the hidden service to forward connection requests from the
        clients. The base32 encoding of the hash of the hidden service's public 
        key (\textit{onion address}) is then communicated to the clients by conventional
        means (blog post, e-mail, etc.).
\end{enumerate}        

When a client decides to connect to the hidden service, he:
\begin{enumerate}
  \item Determines the list of the responsible HS directories using the onion address and
        downloads the HS descriptor.
  \item Chooses a rendezvous point at random.
  \item Communicates the ID of the rendezvous point to the hidden service's introduction points
        which then forward it to the hidden service.
\end{enumerate}
When the hidden service receives the ID of the rendezvous point, it establishes a connection to it
and the data transfer between the service and the client can start.
All communications between the client and the rendezvouz point, between the service and the rendezvous point
and between the service and the introduction points are established over three-hop circuits. This hides the
location of the hidden service and its clients both from each other and from external observer.

The hidden service or a client can determine the fingerprints of the responsible directories
as follows. They first take all Tor relays which have \code{HSDir} flag in the consensus and
sort their fingerprints in lexicographical order. Second, they compute the descriptor IDs of
the hidden service which is the SHA-1 hash of a value composed of the following items\footnote{A hidden service
may also decide to use a secret key (somewhat misleadingly called descriptor-cookie), but for hidden services which
are meant to be accessed by everybody it is not relevant.}: public key
of the hidden service, current day, and replica (which can be 0 or 1). The exact expression for
the ID is of little importance here, the only important things are a) the ID changes every 24 hours,
b) there are two replicas of the ID.
Third they find the place in the sorted list of the fingerprints
for the computed ID and take
the next three relays' fingerprints (thus having 6 fingerprints it total since there are two replicas).

%When Tor receives a new stream from an application it 
%make connection to the Internet through Tor by using it
%Tor's timeout is 2 minutes. For the first two tries Tor waits for 10 secodns and
%then tries another circuit. For tires 3+ it waits for 15 seconds before trying
%a new circuit. Tor Exit nodes can also send a ``TTL Expired'' messages.

%If the client asks to connect to a hostname (not an IP address) Exit nodes normally
%send an END cell with host ``unreachable'' code. The messages arrive fast (apprx. 5 seconds).
%For every such message Tor tries another
%circuit for the stream. After three such messages, Tor gives up the stream.

%\subsection{Banning Tor exit nodes}
%Authors in [] used exploited the Bitcoin built-in reputation-based
%DoS protection to force all Bitcoin servers to ban all Tor Exit nodes.
%In this section we exploit the DoS protection in the similar way however
%we achieve completely different results.
%In the further text we discuss Tor, but the same method
%applies to other anonymity services with minor modifications.

%

\section{Getting in the middle}
\label{sec:getinthemiddle}
By exploiting Bitcoin's anti-DoS protection a low-resource attacker can
force users which decide to connect to the Bitcoin network through
Tor to connect exclusively through her Tor Exit nodes or to her Bitcoin peers, totally isolating the client from the rest of the Bitcoin P2P network.
This means that combining Tor with Bitcoin may have serious security implications for the users:
1) they are exposed to attacks in which an attacker controls  which Bitcoin blocks and transactions the users is aware of;
2) they do not get the expected level of anonymity.

The main building blocks of the attack are: Bitcoin's reputation-based
anti-Dos protection, Tor's stream management policy, the fact that
connections between Bitcoin peers are not authenticated.
Authors in \cite{deanonbitcoin} exploited the Bitcoin's reputation-based
DoS protection to force all Bitcoin servers to ban all Tor Exit nodes.
In this section we exploit the DoS protection, however
 we noticed that instead of just baning Bitcoin clients from using Tor the attacker might achieve much smarter results.
%
%In this section we describe a way for a low-resource attacker to get
%in the middle between Tor users and Bitcoin network.
%One way to achieve this result is to run a number of high-bandwidth Exit Tor relays
%and get control over users which were unlucky to choose the attacker's Exit nodes.
%However this method has several drawbacks:
%\begin{enumerate}
%  \item While some out of the 8 user's connections can exit at the attacker's node,
%        some of them can go through other exit nodes. In this case the
%        attacker will be able to mount traffic correlation attack but will
%        not gain full control over user's view of the network.
%  \item Expensive or very limited number of users.
%        Though previous research showed that an attacker can lie about its bandwidth
%        in the Tor network.
%\end{enumerate}
%
%One can do better by exploiting the Bitcoin's reputation system.
The attack consists of four steps:
\begin{itemize}
  \item Inject a number of Bitcoin peers to the network. Note that though
        Bitcoin allows only one peer per IP address, it does not require
        high bandwidth. IP addresses can be obtained relatively cheap and on per-hour basis.
  \item Periodically advertise the newly injected peers in the network so that they are
        included into the maximum possible number of buckets at the client side.
  \item Inject some number of meduim-bandwidth Tor Exit relays. Even a small
        fraction of the Exit bandwidth would be enough for the attacker as will
	    be shown later.
  \item Make non-attacker's Bitcoin peers ban non-attacker's Tor Exit nodes.
\end{itemize}

We now explain each step of the attack in more detail. See section~\ref{sec:delays} for attack 
parameter estimation.

\subsection{Injecting Bitcoin peers}
This step is rather straightforward. In order to comply with Bitcoin's limitation
``one peer per IP address'', the attacker should obtain a large number of IP addresses.
The easiest way would be to rent IP addresses
on per hour basis. The market value is 1 cents per hour per IP address.
The important note is that the obtained  IP addresses
will not be involved in any abusive activity (like sending spam or DoS attacks) which
makes this part of the attack undetectable.

\subsection{Advertising malicous peers}
The attacker is interested in that her Bitcon peers are chosen by Bitcoin clients as
frequently as possible. In order to increase by factor four the chances for her peers to be included
into ``tried'' buckets, the attacker should advertise the addresses of her
peers as frequently as possible. This mechanism would allow the attacker to inject less 
malicious peers. Note also that address advertisement is not logged by default 
and thus requires special monitoring to be noticed. 

\subsection{Injecting Tor Exit nodes}
During this step the attacker runs a number of Exit Tor nodes. In order to get an Exit
flag from Tor authorities, an attacker's Exit node should allow outgoing connection
to any two ports out of ports 80, 443, or 6667. Such an open Exit policy might not be
what a stealthy attacker wants. Fortunately for the attacker she can provide incorrect information
about her exit policy in her descriptor and thus have an Exit flag while
in reality providing access to port 8333 only. The attacker can do even better,
and  dynamically change the exit policy of her relays so that only connections
to specific Bitcoin peers are allowed.
We implemented this part of the attack: while the Tor consensus indicated that our relays allowed exiting
on ports 80, 443, and 8333 for any IP address, the real exit policy of our relays
was accepting port 8333 for a couple of IP addresses\footnote{We
also allowed exiting to 38.229.70.2 and 38.229.72.16 on any port which are IP addresses
used by Tor bandwidth scanners.}.

\subsection{Banning Tor Exit nodes}
In this phase, the attacker exploits the  built-in Bitcoin anti-Dos protection.
The attacker chooses a non-attacker's Bitcoin peer and a non-attacker's Tor Exit,
builds a circuit through this Exit node and sends a malformed message to the chosen
Bitcoin peer (e.g. a malformed coinbase
transaction which is 60 bytes in size and which causes the immediate ban for 24 hours).
As soon as the Bitcoin peer receives such message it analyses the sender's IP address
which obviously belongs to the Tor Exit node chosen by the attacker. The Bitcoin peer
then marks this IP address as misbehaving for 24 hours. If a legitimate client then
tries to connect to the same Bitcoin peer over the banned Exit node, his connection will
be rejected.
The attacker repeats this step for all non-attackers's Bitcoin peers and each 
non-attacker's Tor Exit node.
This results in that a legitimate Bitcoin user is only able to connect to Bitcoin over Tor
if he chooses either one of the attacker's peers or establishes a circuit through
an attacker's Exit node.
We validated this part of the attack by forcing about 7500 running 
Bitcoin peers to ban our Exit node. To do this we implemented a rudimentary Bitcoin
client which is capable of sending different custom-built Bitcoin messages.

\subsection{Defeating onion peers}
Bitcoin peers can be made reachable as Tor hidden services. Banning Tor Exit nodes will
obviously not prevent Bitcoin clients from connecting to such peers. Nonetheless our
observations show that this case can also be defeated by the attacker.

First the current design of Tor Hidden Services allows a low-resource attacker to DoS a
hidden service of her choice~\cite{trawl} (this technique is called \textit{black-holing} of hidden services). Before a client can contact a hidden service
he needs to download the corresponding descriptor from one of the six responsible hidden
service directories. These directories are chosen from the whole set of Tor relays in
a deterministic way based on the onion address and current day (see section~\ref{sssec:torhs}).
The attacker needs to inject six malicious relays that would become responsible directories.
In other words she needs to find the right public keys
with fingerprints which would be in-between the descriptor IDs of the hidden service and
the fingerprint of the currently first responsible hidden service directory. 
Authors in~\cite{trawl} show that computationally it is  easy to do. It can become a problem though
for a large number of hidden services: for each hidden service the attacker needs to run
at least 6 Tor relays for at least 25 hours, 2 relays per IP address.

Fortunately for the attacker the fraction of Bitcoin peers available as Tor hidden services
is quite small. During August 2014 we queried address databases of reachable Bitcoin
peers~\cite{bitnodes} and among 1,153,586  unique addresses (port numbers were ignored),
only 228 were OnionCat addresses and only 39 of them were actually online; in November
2014 we repeated the experiment and among 737,314 unique addresses 252 were OnionCat
addresses and 46 were online (see Appendix~A
for the two lists of these Bitcoin onion addresses). This 
results in (1) a very small probability for a client to choose a peer available as a hidden
service; (2) this makes black-holing of existing Bitcoin hidden services practical.

Second, the attacker can at almost no cost inject a large number of Bitcoin peers available
as Tor hidden services. It requires running only one \code{bitcoind} instance and binding
it with as many onion addresses as needed. Thus users will more likely connect to attacker controlled ``onion'' peers.

Third, as was described in section \ref{sssec:bcboot}, when running Bitcoin without Tor, onion
addresses received from peers are silently dropped. Thus one can only obtain OnionCat addresses
by either connecting to an IPv4- or IPv6-reachable peers through a proxy\footnote{Not necessarily Tor.}
or by specifying an onion address in the command line.

\section{User fingerprinting}
\label{sec:fingerprinting}
In this section we describe a technique which can be used to fingerprint
Bitcoin users by setting an "address cookie" on the user's computer. 
The cookie can be set and checked even when the user connects to the Bitcoin network through Tor or 
through a chain of proxies. It can be used to correlate different
transactions of the same user even across different sessions (i.e. after
his computer was rebooted). If the user decides later to send a non-sensitive
transaction without Tor, his fingerprint can be correlated to his IP address,
thus deanonymizing all his transactions sent previously through Tor.
The fingerprinting technique is based on the Bitcoin's peer discovery mechanism.
More specifically on that a Bitcoin peer stores addresses received
unsolicitedly from other peers
and on that his database can be queried.

As was described in section~\ref{sssec:bcboot}
whenever a peer receives an unsolicited \code{ADDR} message, it
stores the addresses from this message in his local database.
The attacker can use this fact as follows. When a client connects
to an attacker's peer, the peer sends him a
unique combination of possibly fake addresses (\textit{address cookie}) or fingerprint 
(we will use these two terms interchangably below).
Unique non-existent peer addresses work best, however a more sophisticated and more stealthy adversary may 
use existing bitcoin peer addresses as well (exploiting the combinatorics of the coupon collector problem). 
The client stores these addresses and the next time he connects
to (another) malicious
peer, the peer queries his address database. If the fingerprint addresses are
present in the set of retrieved addresses, the attacker identifies the user.

Consider a user $C$ and a set of Bitcoin servers $E_1,...,E_k$ controlled by an
attacker. Assume that one of the attacker's servers $E_l$ is among the user's
entry nodes. The attacker executes the following steps:
\begin{enumerate}
  \item Send a number of \code{GETADDR} messages to the user. The user should reply
        with \code{ADDR} messages.
  \item Check the received from the client addresses if they already contain a
        fingerprint. If the user already has a fingerprint, stop. Otherwise go to
        the next step.
  \item Generate a unique combination of $N$ fake addresses $FP$ and send them in
        an \code{ADDR} message to the client. The \code{ADDR} message should contain
        at least 11 addresses so that it is not forwarded by the client. 
        If $N$ is less than 11, pad the message with $11-N$ legitimate\footnote{By
        legitimate we mean that there are some Bitcoin servers
        running at these addresses.} addresses.
  \item If the user connects to the Bitcoin network directly (i.e. without Tor),
        store the correspondence between the client's IP address and his fingerprint
        as a tuple $(FP,IP_C)$. If the user connect through Tor save him as
        $(FP,\text{NIL})$.
\end{enumerate}

There is a detail of the Bitcoin protocol which an attacker should take into account.
As was described in subsection \ref{sec:background}-A1, when a client connects
to the Bitcoin network over Tor, he will accept and store in his database
OnionCat addresses only (thus ignoring IPv4 addresses).
It means that in case of Tor,  the fingerprint generated by the attacker should consist
of OnionCat addresses only.
On the other hand when a client connects to the network directly, he will ignore
non-IPv4/IPv6 addresses. Hence an attacker should generate a fingerprint
consisting of IPv4 addresses only. This results that an attacker needs
to store 2 different types of cookies: OnionCat and IPv4.
At the same time, a client does not limit the types of addresses he sends as
a reply to a \code{GETADDR} message. This means that once a cookie was set it can
be queried both over Tor and directly.

\subsection{Stability of a Cookie}
According to the Bitcoin core~\cite{bccore} source code, at the startup when a client establishes outgoing
connections he sends \code{GETADDR} messages, and gets back a set of addresses (typically
2,500, the maximum possible number per \code{GETADDR} request). Given 8 outgoing connection,
the client will receive up to 20,000 non-unique addresses. These addresses can potentially
overwrite the address cookie previously set by an attacker. Below we will try to estimate how
this affects the stability of the cookie. Assume that an attacker managed to set
an address cookie on a user's computer and disconnected (ex. client ended the session). 
The client then establishes a new session sometime later.

First note that if the users reconnects to Bitcoin over Tor and if the attacker has mounted the attack from
section~\ref{sec:getinthemiddle}, he controls all user's traffic and the cookie is preserved.
Let us now describe what happens if the client decides to connect to the Bitcoin network directly.

When a client receives an address $IP_{in}$ he
first checks if it is already contained in his database. If yes, he does nothing (thus the cookie
is not damaged). In case it is a new IP address  the client executes the following procedure.
He computes the bucket number (see section~\ref{ssec:buckets}) based
on the peer which sent the address and the address itself.
If this bucket contains a ``terrible''\footnote{An address is called terrible if any of the following holds:
1) its timestamp is 1 month old or more than 10 minutes in the future; 2) 3 consecutive connections
to this address failed.} address $IP_{terrible}$, it is replaced by $IP_{in}$. Otherwise 4 random
addresses are chosen from the bucket and the one with the oldest timestamp is replaced by $IP_{in}$.

In other words, in order for the incoming address $IP_{in}$ to replace a cookie
address $IP_{cookie}$\footnote{A cookies consists of several IP address, but in order to make
the explanation simpler, we use just one address here.} the following conditions should hold:
\begin{enumerate}
  \item $IP_{in}$ should not be in the user's database;
  \item $IP_{in}$ should belong to the same bucket $B$ as $IP_{cookie}$ and
        there should be no ``terrible'' addresses in $B$;
  \item $IP_{cookie}$ should be among the four randomly chosen addresses, and
        its timestamp should be the oldest.
\end{enumerate}
\noindent These conditions as we will see below make the attacker's cookie quite stable for many hours (this also depends on the number of user sessions since at each startup the address database is refreshed).

In order to estimate the probability that a cookie address set by the attacker is preserved we
conducted the following experiment. In November 2014 we queried running Bitcoin servers by
sending them \code{GETADDR} messages. We received 4,941,815  address-timestamp pairs.
Only 303,049 of the addresses were unique. This can be interpreted as that only about 6\% of
the addresses received by a client will not be already contained in his database
(if the client re-connects immediately).

As the second step, we looked at the timestamps distribution of the non-unique address set.
This distribution can serve as approximation of the
distribution of address timestamps of a client's database.
The results are shown in Table~\ref{tab:timestamps}: 89\% of addresses
had a timestamp more than 3 hours in the past. Taking into account conditions stated
above, it almost guarantees that the attacker's cookie will not be damaged within the
first 3 hours. For 45\% of addresses the timestamp was older than 10 hours (which is the
duration of a working day); 9\% of addresses were older than 1 week.

\begin{table}[h!]
\centering
{\small
\begin{tabular}{|c|c|}
\hline
Address age, hours & 1-CDF \\ \hline \hline
3  & 89\% \\ \hline
5  & 77\% \\ \hline
10  & 45\% \\ \hline
15  & 28\% \\ \hline
24  & 19\% \\ \hline
36  & 15\% \\ \hline
48  & 13\% \\ \hline
72 (3 days)  & 12\% \\ \hline
168 (1 week)  & 9\% \\ \hline
\end{tabular}
}
\caption{Complementary Cumulative distribution function for addresses timestamps}
\label{tab:timestamps}
\end{table}

The results above could be summarized as follows: (1) there is a high chance that an
address received by a client will already be contained in his database, which keeps
the cookie intact; (2) if a cookie IP address is among the 4 nominees for erasing,
it is likely that its timestamp will be fresher than that of at least one of other nominees (and thus
will not be erased).

Finally we conducted the following experiment. We set a cookie consisting of 100 IPv4 addreses
and monitored how stable this cookie was across different sessions. Table~\ref{tab:cookiecheck} shows
the decay rate of the number of cookie addresses over time and sessions. Note that by session we mean
that the client switched off Bitcoin software and switched it on again, which forces him to make 8 new outgoing
connections and retrieve up to 20,000 addresses.

\begin{table}[h!]
\centering
{\small
\begin{tabular}{|c|c|c|}
\hline
Session number & Time since start, hours & Remaining addresses\\ \hline \hline
1  & 0    & 100 \\ \hline
2  & 0.5  & 100\\ \hline
3  & 1    & 100 \\ \hline
4  & 1.5  & 100 \\ \hline
5  & 2    & 100 \\ \hline
6  & 2.5  & 100 \\ \hline
7  & 3    & 98 \\ \hline
8  & 3.5  & 92 \\ \hline
9  & 5.5  & 50 \\ \hline
10  & 8   & 36 \\ \hline
\end{tabular}
}
\caption{Address cookie decay rate (example)}
\label{tab:cookiecheck}
\end{table}

\noindent The experiment shows that even after 10 sessions (i.e. after
reception of about 200,000 non-unique IP addresses) and 8 hours, one third
of the fingerprint remained in the user's database (thus it will be possible to identify the client). Note that sessions
9 and 10 took 2 and 2.5 hours. On the average an attacker will
need about 90 peers (given that at the time of writing there are about 7,000 Bitcoin servers) to become one the client's
entry nodes during any of these 10 sessions and update the fingerprint.
Running this number of peers will cost the attacker less than 650 USD per month
(see section~\ref{sec:costsandrate}).

In another experiment we checked that in the case of two sessions 
with 10 hours between sessions, our client kept
76\% of the initial fingerprint addresses, and in the case of 
24 hours between two sessions 55\% of the initial fingerprint
were kept (which again allows user's identification).
In order to carry out the experiments from this section we built our own
rudimentary Bitcoin client which is able to connect/accept connections
to/from Bitcoin peers
and is capable of sending/receiving different Bitcoin messages on demand.
We used this client as a malicious Bitcoin server which sets new address
cookies and checks previously set cookies. In order to simulate a user
we used the official Bitcoin core software
(developed by the Bitcoin project)~\cite{bccore}.
The attack from this section was experimentally verified by
tracking our own clients in the real Bitcoin and Tor networks.

\subsection{Cookie extraction}
The remaining question is how many \code{GETADDR} messages an attacker needs to send
to the client to learn that the database of this client contains a cookie.
According to~\cite{deanonbitcoin}, section 9.2 it can be up to 80 messages to retrieve the full collection
of client's addresses.
However in practice we will not need to collect all the addresses
in a fingerprint, which significantly reduces the number of requests.
About eight \code{GETADDR} messages would be sufficient to retrieve about 90\% of the cookie addresses. This shows
that the cookie can be checked without raising suspicion.

\subsection{Deanonymization of Bitcoin over Tor users}
Consider the following case. A client uses the same computer for sending both
benign Bitcoin transactions and sensitive transaction.
For benign transactions the user connects to Bitcoin directly,
but for sensitive transactions he forwards his traffic through a chain of Tor relays
or VPNs.
If an attacker implements the attack described in section~\ref{sec:getinthemiddle},
all client's sensitive transactions
with high probability will go through attacker's controlled nodes
which will allow her to fingerprint the user and record his transactions. 

When the client later connects to the Bitcoin network directly to send benign transactions,
he will with some probability choose an entry node controlled by the attacker (in
section~\ref{sec:portspoof} we show how to increase this probability). Once it happens,
the attacker can query the client for the fingerprint and thus correlate his sensitive
transactions with his IP address. Note that even if the attacker is not 
implementing the complete man-in-the-middle attack on Tor, 
but just injects Sybil peers and Sybil hidden services he will be able
to link many sensitive transactions to the real IP addresses of the users.

\subsection{Linking different Tor sessions}
In the case, when a client uses a separate computer (or
Bitcoin data folder\footnote{A bitcoin data folder
is a directory where Bitcoin clients store their wallets
and dump IP address databases between restarts.}), the attacker will not be able 
to learn his IP address. However, the attacker will still be able to link different
transactions of the same user (remember that if a client sends a transaction through Tor
the attacker can be certain that it was generated by this client). This can be done even across different
sessions (computer restarts). This will in turn allow the attacker to correlate
different Bitcoin addresses completely unrelated via transaction graph analysis.

\subsection{Domino Effect}

Tor multiplexes different streams of the same user over the same circuits.
This means that if the source of one stream in the circuit is revealed by
the fingerprinting attack, all other streams will also be deanonymized.
Specifically, it is likely that a user who sends a sensitive Bitcoin transaction through Tor, will also browse a darkweb site.
Similar result was also noted in~\cite{bittorrenttor} but in relation to
Bittorrent over Tor privacy issues.
\section{Low-resource sybil attacks on Bitcoin}
\label{sec:portspoof}

In the previous section we mentioned that a client needs to connect directly to
one of the attacker's nodes in order to reveal his IP address
so that an attacker can deanonymize his previous transactions done over Tor.
Bitcoin as a peer-to-peer network is vulnerable to Sybil attacks and just operating
many Bitcoin servers means that a client will sooner or later choose an entry nodes controlled
by the attacker (i.e. in some number of sessions). However running too many servers can be costly 
(see section~\ref{sec:costsandrate} for attack cost estimation).
Fortunately for the attacker there are a couple of ways to prevent Bitcoin clients from using
non-attacker's Bitcoin servers (and choose an attacker's one instead).

\subsection{Exhausting connections limit}
As described in section~\ref{sec:background}, by default a Bitcoin server accepts
up to 117 connections. Once this limit is reached all new incoming connections are
dropped. At the same time a Bitcoin server neither checks if some of these connections
come  from the same IP address\footnote{One explanation is that if clients are behind the
same NAT they will share the same IP address.}, nor forces client to provide a proof-of-work.
As a result a low-resource attacker can establish many connections to all but his Bitcoin servers\footnote{The list of all running Bitcoin servers can be
obtained from e.g.~\cite{bitnodes}.}
and occupy all free connection slots.
If a client connects directly to a Bitcoin server connections slots of which are occupied,
the connection will be dropped immediately, thus the client will soon end up connecting to
a malicious peer. This straightforward attack has been known in the Bitcoin community.

\subsection{Port poisoning attack}
A less effective but much stealthier new attack exploits the following fact.
Peer addresses are of the following form (IP,PORT). However when a client
decides if to add a received address to the database, he does not take the port
number into account. For example
assume a client receives an address $(IP_0, PORT_1)$ and there is already
an entry in the client's database $(IP_0, PORT_0)$. In such case the client
will keep $(IP_0, PORT_0)$ and will not store $(IP_0, PORT_1)$.

The attacker can use this fact to flood with clients with addresses of
legitimate Bitcoin servers but wrong port numbers. If the attacker is the first
to send such addresses, the client will not be able to connect to legitimate nodes.

\section{Estimating client's delays}
\label{sec:delays}
The steps described in section~\ref{sec:getinthemiddle} imply that once a client
decides to use Bitcoin network over Tor, he will only be able to do this
by choosing either one of the attacker's Exit nodes or one of the attacker's Bitcoin
peers. However for the attack to be practical a user should not
experience significant increases in connection
delays. Otherwise the user will just give up connecting and decide that Tor-Bitcoin bundle is malfunctioning. 
In this section we estimate the number of Bitcoin peers and the amount of bandwidth of 
Tor Exit relays which the attacker needs to inject, so that the attack does
not degrade the user's experience.

\begin{figure}[h]
\begin{center}
\includegraphics[scale=0.4]{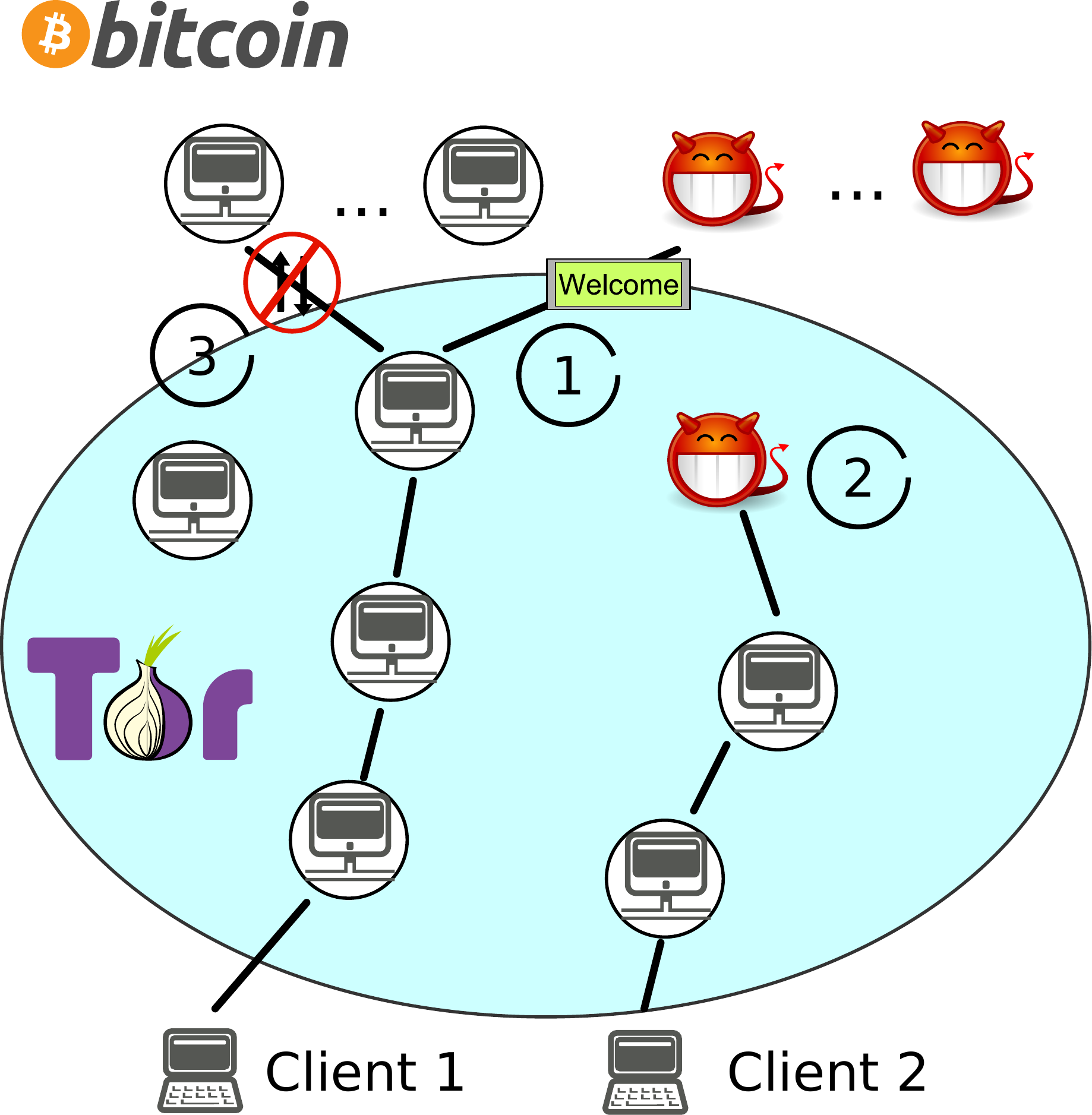}
\caption{Client's state after the main steps of the attack}
\label{fig:attacksteps}
\end{center}
\end{figure}

Once the attacker completes the steps described in the previous section, 
for each user connecting to the Bitcoin
network through Tor there are several possibilities (see Fig.~\ref{fig:attacksteps}).
\begin{enumerate}
  \item The user chooses one of the attacker's Bitcoin peers. The attacker
        does nothing in this case: the attacker
        automatically gains control over the information forwarded to the user.
  \item The user chooses one of the attackers Exit nodes. The attacker can use
        the fact that Bitcoin connections are not encrypted and not authenticated and
	     redirect the client's request to Bitcoin peers under her control.
  \item The user chooses a non-attacker's Exit relay and a running non-attacker's
        Bitcoin peer. In this case, due to the ban the user's connections
	    will be rejected. And the user will try to connect to a different
	    Bitcoin peer.
  \item The user chooses a non-attacker's Exit relay and a non-attacker's
        Bitcoin peer which went offline\footnote{Or never really existed: Bitcoin
        allows storing fake addresses in client addresses database.}.
        In this case the Bitcoin client will wait until the connection
        times out which can be up to two minutes (see section~\ref{sssec:tortimeoutes}). This delay 
        on the surface looks like taking prohibitively long time. However since during these 
        two minutes Tor rebuilds new circuits every 10-15 seconds, trying new Exits at random, it                actually makes the attacker's life much easier. It increases the chances that malicious                  Exit relay will be chosen.
\end{enumerate}

\subsection{Handling unreachable Bitcoin peers}
Before estimating the delays we consider case 4 in more detail.
Our experiments show that for a Bitcoin client which was already used several
times prior to the connection over Tor, the addresses database contains
10,000 -- 15,000 addresses and the fraction of unreachable
Bitcoin peers among them is between 2/3 and 3/4. Abundance of unreachable addresses  means that case 4 is the most frequent scenario for the client. Consider a client which chose an unreachable Bitcoin server and a non-attacker's
Exit node. 

The Exit relay can send either:\footnote{This is based on
the Tor source code analysis and monitoring a running Tor instance.}
\begin{enumerate}
  \item An END cell with a ``timeout'' error code.
        In case of a ``timeout'' message, Tor sends a ``TTL expired''
        SOCKS error message
        to the Bitcoin application which then tries another Bitcoin peer.
  \item An END cell with ``resolve failed' error code\footnote{We observed
        this behaviour not only for hostnames but also for IP addresses.}.
        In case of ``resolve'' fail message, Tor drops the current circuit and
        tries to connect to the unreachable Bitcoin peer through a different
        Exit node. After 3 failed resolves, Tor gives up and sends a ``Host unreachable''
        SOCKS error code, which also results in Bitcoin trying a different peer.
\end{enumerate}
        
The third and the most common option is that the exit relay
will not send any cell at all during 10-15 seconds.
As was described in the Background section that in case the Exit node does not send any reply
within 10 or 15 seconds (depending on the number of failed tries) along the
circuit attached to the stream, Tor drops the current circuit and attaches
the stream to another circuit (or to a newly built one if no suitable circuits
exist). In case Tor cannot establish connections during 125 seconds, it
gives up and notifies Bitcoin client by sending a ``General failure''
SOCKS error message. Bitcoin client then tries another peer.

\subsection{Estimating delays}

The facts that a) Tor tries several different circuits while trying to connect
to unreachable peers and b) the fraction of unreachable peers in the
client's database is very large, significantly increases the chances that a
malicious Exit node is chosen. The attacker only needs this to happen once, since
afterwards all connections to the other Bitcoin peers will be established through
this Tor circuit; Bitcoin client will work even with one connection.
On the other side, unreachable nodes increase the delay before the user
establishes its first connection. This delay depends on the number of
attacker's Bitcoin peers and on how often the user chooses new circuits.

\begin{figure}[h]
\begin{center}
\includegraphics[scale=0.6]{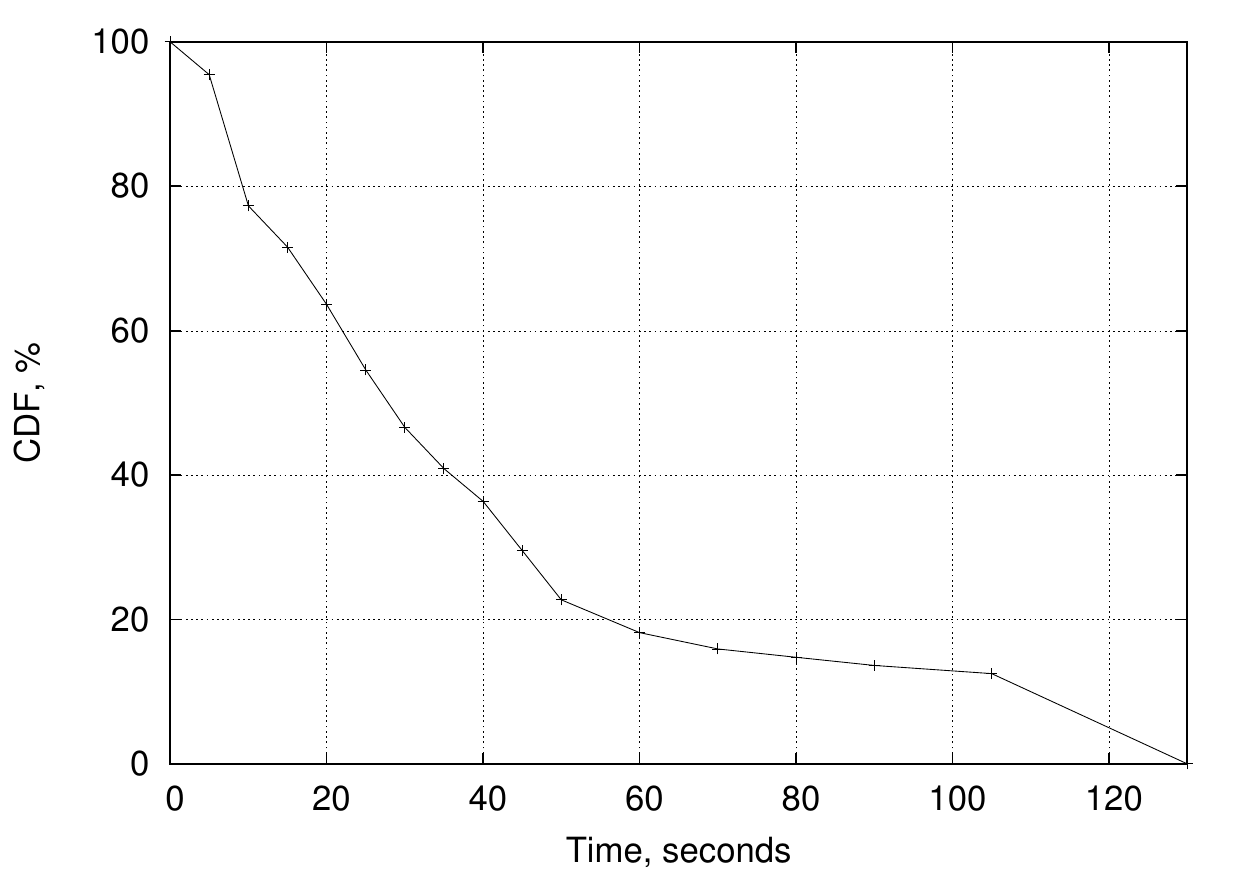}
\caption{Time spent connecting to an unreachable node}
\label{fig:failroundcdf}
\end{center}
\end{figure}

In order to estimate the later, we carried out the following experiment.
We were running a Bitcoin client over Tor and for each connection to
an unreachable Bitcoin client we were measuring the duration of the
attempt and the number of new circuits (and hence different Exit nodes).
The cumulative distribution function of the amount of time a Bitcoin
client spends trying to connect to an unreachable node is shown in
Fig.~\ref{fig:failroundcdf}. On the average a Bitcoin peer spends 39.6 seconds
trying to connect to an unreachable peer and tries to establish
a new circuit (and hence a different Exit node) every 8.6 seconds.
This results in 4.6 circuits per unreachable peer.
%In the first set of experiments we took into account the fact that
%every second connection was made to an unreachable hostname (one of the six
%hostnames which are used in DNS-seed bootstrapping).

\begin{figure}[h]
\begin{center}
\includegraphics[scale=1.5]{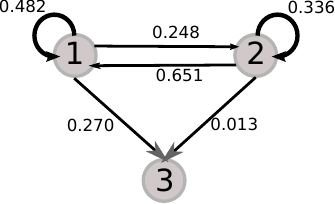}
\caption{Markov chain with probabilities for 400K of Exit capacity and 100 malicious Bitcion peers. The client spends about 0.5 seconds in State 2 and about 40 seconds in State 1}
\label{fig:markov}
\end{center}
\end{figure}

We now estimate how long it will take a user on average to establish his
first connection to the Bitcoin network. This delay obviously depends on
the number of the attacker's Bitcoin peers and the amount of bandwidth
of her Tor Exit relays. We adopt a simple discrete time absorbing Markov chain model
with only three states (see Fig~\ref{fig:markov}): 
\begin{itemize}
\item  State 1: the Bitcoin client tries to connect to an unreachable
peer; 
\item State 2: the Bitcoin client tries to connect to a reachable Bitcoin peer banned
by the attacker; \item State 3: Bitcoin peer tries to connect to an attacker's Bitcoin peer
or chooses an attacker's Tor Exit node. State 3 is absorbing state, once it is
reached, the user thinks that he connected to the Bitcoin network (while he is
now controlled by the attacker). 
\end{itemize}

\begin{figure}[h]
\begin{center}
\includegraphics[scale=0.6]{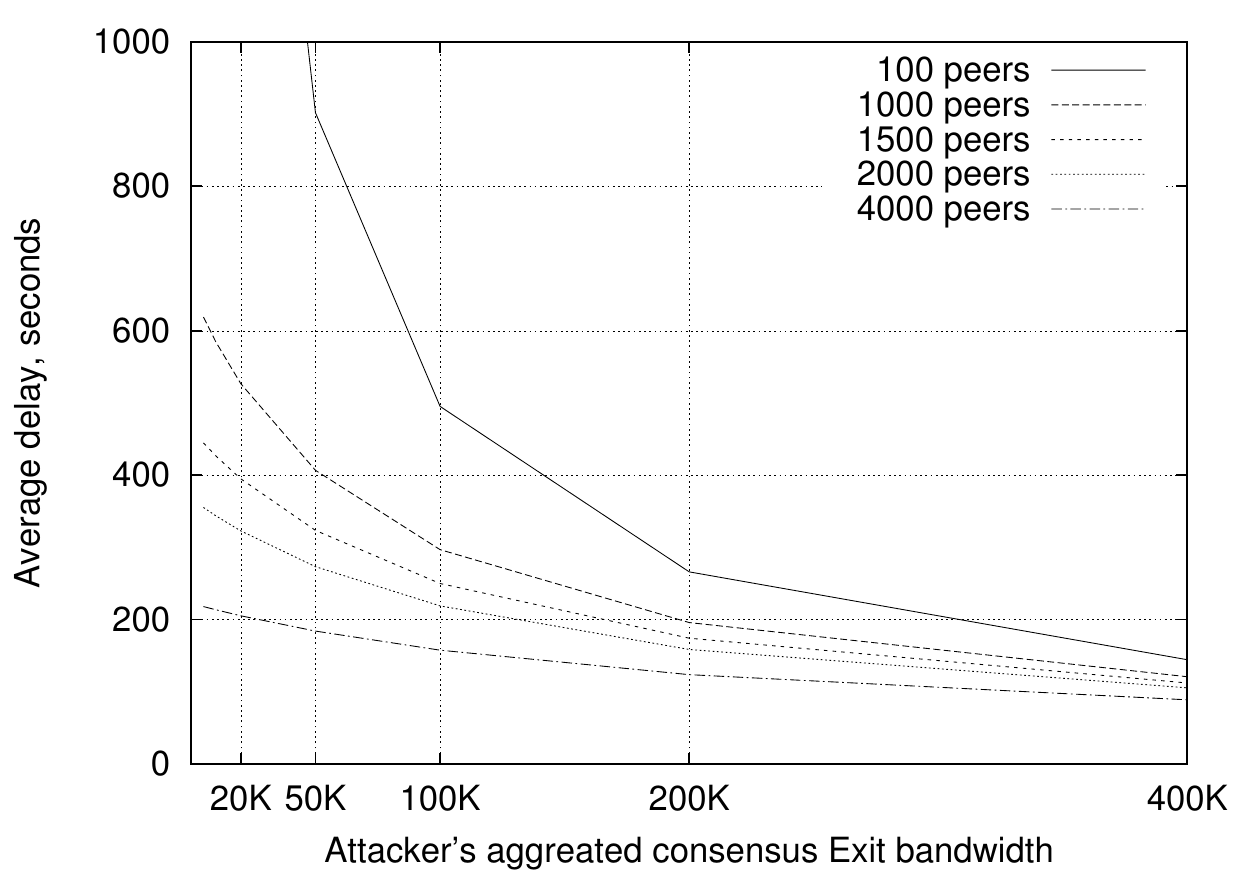}
\caption{Average time before the first connection}
\label{fig:userdelays}
\end{center}
\end{figure}

\noindent After composing the fundamental matrix for
our Markov chain, we find the average number of steps in two non-absorbing states.
Taking into account the average amount of time spent by the user in each of the
states (we use our experimental data here), we find the average time before
the absorbing state. We compute this time for different number of Bitcoin peers
and Tor Exit relay bandwidth controlled by the attacker.
The results are presented in Fig~\ref{fig:userdelays}. We have taken a concervative estimate that the fraction
of unreachable Bitcoin peers in the client's database is 2/3 = 66\%, also the client spends only about 0.5 seconds in State 2 and about 40 seconds in State 1.
%would be nice to measure more accurately what happens on average 2/3 or 3/4?

\noindent Fig~\ref{fig:userdelays} shows that an attacker having
100,000 of consensus Exit bandwidth and 1000 Bitcoin peers is able to carry
out the attack while keeping the average delay below 5 minutes. For example attacker controlling a small botnet can afford that many peers (she will need 1000 peers with public IPs or supporting  UPnP protocol).
An attacker having consensus weight of 400,000 and very few peers
can decrease the average delay to about two minutes. Such a bandwidth
is achievable by
an economy level attacker as will be shown in section~\ref{sec:costsandrate}.

The line corresponding to 4000 attacker's Bitcoin peers
in Fig.~\ref{fig:userdelays} is not as unrealistic as it may seem.
Recall (see section \ref{ssec:buckets}) that each Bitcoin peer address
can go to up to 4 ``new'' buckets at the client's side. This can be used
by a persistent attacker to increase the choice probability for her
peers by a factor 4 (in the best case) which means an attacker can have significantly less
than 4000 peers.

\subsection{Clients with empty addresses cache}
As was pointed in section~\ref{sssec:bcboot}, all IPv4 and IPv6 addresses
received from DNS-oneshots are dropped by a Bitcoin client if Tor is used.
If the addresses database of a client is empty and all the
seed nodes are banned, the client can connect to hidden services only.
This is a limitation of our approach.

\section{Opportunistic deanonymisation and traffic correlation}
For a traffic confirmation attack an attacker needs to control both
ends of a users's communication and the fact that the attacker now
controls one end of the communication significantly increases the success
rate of the attack.
The attacker sends a traffic signature down the circuit in hope that
some users chose the attacker's Guard nodes. Even if it is not the case,
the attacker can try to reveal the Guard nodes of users by dropping their
circuits and forcing them to reconnect (thus choose another circuit which
would contain one of the attacker's middle nodes).

Attacker would be also in position to perform
time correlation attacks in case the user is accessing Tor Hidden services and making payments with Bitcoin over Tor on them. For this it would be sufficient for the attacker to get control of the HSDir servers responsible for the onion addresses of the relevant hidden service.

\subsection{Possibility of double spending and "virtual reality" }
The attacker can defer transactions and blocks. In case of blocks we can send dead
forks. In collusion with a powerful mining pool (for example 10-20\% of total Bitcoin mining capacity) the attacker can create fake blocks. This enables double spend, however to make this relevant the amount should exceed what such miner would be able to mine in the real Bitcoin network. Also complete alternative Bitcoin reality for all the users who access Bitcoin solely through Tor is possible. This however would come at a cost of 5-10 times slower confirmations, which after some time can be detected by the wallet software. 

\section{Attack Costs}
\label{sec:costsandrate}
\subsection{Tor Exit nodes}
During July 2014 we were running a non-Exit Tor relay for 30 USD per month.
We set the bandwidth limit of the relay to 5 MB/s which resulted in traffic
of less than 15GB per hour. The consensus bandwidth of
this relay fluctuated between 5,000 and 10,000 units\footnote{A unit roughly
corresponds to 1 KB/s of traffic}. While the total weighted
consensus bandwidth of all exit nodes was about 7 million units, the weighted
consensus bandwidth of relays allowing exiting at port 8333 was about
5.7 million units. Assuming that we could achieve the same consensus bandwidth for
an Exit node this gives the probability of 0.08\%-0.17\% for our relay to be
chosen for Exit position by a user. Given that 10 TB of traffic is included
into the server's price and one has to pay 2 EUR per additional 1 TB,
it would cost an attacker 360 USD to have 180 TB of traffic per month.
The corresponding speed is 69 MB/s (69,000 consensus bandwidth units).
By running 6 such relays the attacker can achieve 400K of bandwidth
weight in total for the price below 2500 USD (2160 USD for the traffic and 240 for renting fast servers).

Thus having a consensus weight close
to 400,000 is possible for an economy-level attacker.
The attacker can also decide to play unfair and mount a bandwidth
cheating attack which would allow her to have a high consensus weight
while keeping the budget of the attack even lower.
This is especially possible since Bitcoin traffic by itself is rather
lightweight and high bandwidth would be needed only in order to drive
Tor path selection algorithm towards attacker's nodes. 

\subsection{Bitcoin peers}

The attack described in sections~\ref{sec:getinthemiddle} and
\ref{sec:fingerprinting} suggests the attacker
injects a number Bitcoin peers; at the same time Bitcoin network allows
only one peer per IP address. Thus the attacker is interested in getting as many IP addresses as possible. Currently there are several options.
The cheapest option would be 
to rent IP addresses on hourly basis. The market price for an IP address
is 1 cent per hour~\cite{terremark}. This results in 7200 USD per 1000 IP's 
per month.

From these computations it is clear that attacker would do better by investing in Exit bandwidth rather than running Bitcoin peers (unless she controls a small botnet), and the only limitation for her would be not to become too noticeable. An attacker that has 400K (7\% for port 8333) of Tor Exit capacity would cost about 2500 USD. 

\section{Countermeasures}
\label{sec:countermeasures}
These attacks are very effective due to a feature of Bitcoin which allows an easy ban of Tor Exit nodes from arbitrary Bitcoin peers and due to easy user fingerprinting with the "address cookies". One possible countermeasure against Tor-ban could be to relax the reputation-based DoS protection. For example each Bitcoin peer could have a random variable, which would decide whether to turn ON or OFF the DoS protection mechanism with probability 1/2. As a result the attacker might be able to DoS at most half of the network, but on the other hand he will not be able to ban any relays or VPNs from ALL the Bitcoin peers. 

An obvious countermeasure would be to encrypt and authenticate Bitcoin traffic.  This would prevent even opportunistic man-in-the-middle attacks (i.e. even if the user is unlucky to choose the malicious Exit relay).
Yet another possible countermeasure is to run a set of "Tor-aware" Bitcoin peers which would regularly download Tor consensus and make sure that Bitcoin DoS countermeasures are not applied to servers from the Tor consensus. Such relays however would be vulnerable to DoS via Tor attack.

Finally, Bitcoin developers can maintain and distribute a safe and stable list of onion addresses.
Users which would like to stay anonymous should choose at least one address from this list.
There currently exists a short and not up-to-date list of Bitcoin fallback
onion addresses~\cite{fallbacknodes}.

With regards to the fingerprint attack, a possible countermeasure could
be requiring to perform proof-of-work computation for each
sent \code{GETADDR} message, so that it becomes computationally
expensive for an attacker to query each client. An immediate countermeasure would be to 
remove the cached address database file before each session and to use only trusted hidden-services.

%\IEEEtriggeratref{11}



\bibliographystyle{IEEEtranS}
\bibliography{ndss}

\appendix{A. List of Reachable Bitcoin Onions}
%\section{LIST OF REACHABLE BITCOIN ONIONS}
\label{sec:bconions}

In this Appendix we list 39 Bitcoin onion addresses which we found to be
reachable in August 2014 and 46 onion addresses reachable
in November 2014. In order to get this list we queried reachable for
the time of the experiments Bitcoin peers by sending four \code{GETADDR} messages to each peer.
A Bitcoin peer can reply
to such message by sending back 23\% of its addresses database but not more than 2500 addresses.
A peer can store 20,480 addresses at most which means that sending 4 \code{GETADDR} messages
is not enough to extract the complete peer's database. However we expect that there is a
big overlap between the databases of different peers.
Some of the discovered reachable onion addresses begin or end with meaningful text
like: BTCNET, BITCOIN and belong to Bitcoin developers,
pools or services.

\begin{table}[h]
\centering
\texttt{
{\scriptsize
\begin{tabular}{|l|l|}
\hline
2fvnnvj2hiljjwck.onion:8333  & it2pj4f7657g3rhi.onion:8333 \\ \hline
2zdgmicx7obtivug.onion:8333  & jq57qrkvvyi4a3o2.onion:8333 \\ \hline
3crtkleibhn6qak4.onion:14135 & kjy2eqzk4zwi5zd3.onion:8333 \\ \hline
3lxko7l4245bxhex.onion:8333  & mtzcz5knzjmuclnx.onion:8333 \\ \hline 
4crhf372poejlc44.onion:8333  & nns4r54x3lfbrkq5.onion:8333 \\ \hline
5ghqw4wj6hpgfvdg.onion:8333  & nzsicg2ksmsrxwyz.onion:8333 \\ \hline
5k4vwyy5stro33fb.onion:8333  & pqosrh6wfaucet32.onion:8333 \\ \hline
6fp3i7f2pbie7w7t.onion:8333  & pt2awtcs2ulm75ig.onion:8333 \\ \hline
7iyfdkr72hgtdjoh.onion:8333  & pxl7ytsd2aiydadi.onion:8333 \\ \hline
b6fr7dlbu2kpiysf.onion:8333  & qsxhkpvbmt6akrov.onion:8333 \\ \hline 
bitcoin625tzsusi.onion:8333  & syix2554lvyjluzw.onion:8333 \\ \hline
bitcoinostk4e4re.onion:8333  & t2vapymuu6z55s4d.onion:8333 \\ \hline
btcdatxubbzaw4tj.onion:8333  & td7tgof3imei3fm6.onion:8333 \\ \hline
btcnet3utgzyz2bf.onion:8333  & tfu4kqfhsw5slqp2.onion:8333 \\ \hline
czsbwh4pq4mh3izl.onion:8333  & thfsmmn2jbitcoin.onion:8333 \\ \hline
dqretelgl3kjtzei.onion:8333  & xdnigz4qn5dbbw2t.onion:8333 \\ \hline
e3tn727fywnioxrc.onion:8333  & xij5qyrbosw2pzjm.onion:8333 \\ \hline
evolynhit7shzeet.onion:8333  & zqq6yxxxb7or36br.onion:8333 \\ \hline
gb5ypqt63du3wfhn.onion:8333  & zy3kdqowmrb7xm7h.onion:8333 \\ \hline
hkxy4jpeniuwouiv.onion:8333  &                             \\ \hline
\end{tabular}
}}
\caption{Bitcoin onions, online in August 2014}
\label{tab:onions1}
\end{table}

\begin{table}[h]
\centering
\texttt{
{\scriptsize
\begin{tabular}{|l|l|}
\hline
2xylerfjgat6kf3s.onion:8333 & h2vlpudzphzqxutd.onion:8333 \\ \hline
2zdgmicx7obtivug.onion:8333 & hkxy4jpeniuwouiv.onion:8333 \\ \hline
3ffk7iumtx3cegbi.onion:8333 & iksneq25weneygcj.onion:8333 \\ \hline
3lxko7l4245bxhex.onion:8333 & k22qrck6cetfj655.onion:8333 \\ \hline
4crhf372poejlc44.onion:8333 & kjy2eqzk4zwi5zd3.onion:8333 \\ \hline
5k4vwyy5stro33fb.onion:8333 & lazsruhzupsgpvwm.onion:8333 \\ \hline
6fizop6wctokuxyk.onion:8333 & lfmwsd65ltykrp74.onion:8333 \\ \hline
6fp3i7f2pbie7w7t.onion:8333 & luruc27g24y7ewwi.onion:8333 \\ \hline
7g7j54btiaxhtsiy.onion:8333 & pqosrh6wfaucet32.onion:8333 \\ \hline
7pkm6urc5hlgwlyc.onion:8333 & pt2awtcs2ulm75ig.onion:8333 \\ \hline
b2ykuvob44fn36wo.onion:8333 & pxl7ytsd2aiydadi.onion:8333 \\ \hline
b6fr7dlbu2kpiysf.onion:8333 & qsntokcdbwzmb2i5.onion:8333 \\ \hline
bitcoinostk4e4re.onion:8333 & sbow7bnje2f4gcvt.onion:8333 \\ \hline
bk5ejfe56xakvtkk.onion:8333 & td7tgof3imei3fm6.onion:8333 \\ \hline
btc4xysqsf3mmab4.onion:8333 & tfu4kqfhsw5slqp2.onion:8333 \\ \hline
btcnet3utgzyz2bf.onion:8333 & thfsmmn2jbitcoin.onion:8333 \\ \hline
by4ec3pkia7s7wy2.onion:8333 & ukronionufi6qhtl.onion:8333 \\ \hline
dioq2yg3l5ptgpge.onion:8333 & vqpye2k5rcqvj5mq.onion:8333 \\ \hline
dqretelgl3kjtzei.onion:8333 & wc5nztpe26jrjmoo.onion:8333 \\ \hline
drp4pvejybx2ejdr.onion:8333 & xudkoztdfrsuyyou.onion:8333 \\ \hline
e3tn727fywnioxrc.onion:8333 & z3isvv4llrmv57lh.onion:8333 \\ \hline
evolynhit7shzeet.onion:8333 & zc6fabqhrjwdle3b.onion:8333 \\ \hline
gb5ypqt63du3wfhn.onion:8333 & zy3kdqowmrb7xm7h.onion:8333 \\ \hline
\end{tabular}
}}
\caption{Bitcoin onions, online in November 2014}
\label{tab:onions2}
\end{table}

\end{document}